\documentclass[twocolumn,a4paper,eqsecnum,english,a4paper,showpacs,preprintnumbers,amsmath,amssymb,prb]{revtex4}
\usepackage[latin1]{inputenc}
\usepackage{float}
\usepackage{graphicx}
\usepackage{amssymb}

\makeatletter

\usepackage{float}

\makeatletter

\usepackage{float}

\makeatletter

\usepackage{dcolumn}

\usepackage{bm}

\makeatother

\makeatother

\usepackage{babel}
\makeatother
\begin{document}

\title{Strong coupling theory of spin and orbital excitations 
in Sr$_2$IrO$_4$}

\author{Jun-ichi Igarashi$^{1}$ and and Tatsuya Nagao$^{2}$}

\affiliation{
 $^{1}$Faculty of Science, Ibaraki University, Mito, Ibaraki 310-8512,
Japan\\
$^{2}$Faculty of Engineering, Gunma University, Kiryu, Gunma 376-8515,
Japan
}

\date{\today}

\begin{abstract}

We study the low-lying excitations in 5d transition-metal oxide
Sr$_2$IrO$_4$ on the localized electron picture.
We find that Hund's coupling together with the spin-orbit interaction leads 
to exchange anisotropy which causes the spin wave gap. 
Introducing the isospin operators acting on Kramers' doublet
in Ir atoms, we derive the effective spin Hamiltonian from 
a multi-orbital Hubbard model with the $t_{2g}$ orbitals in the square lattice.
We introduce the Green's functions including the anomalous type
for the boson operators by expanding the spin operators in terms of boson 
operators in the lowest order of $1/S$, and solve the coupled equations 
of motion for those functions. Two modes are found to emerge with slightly 
different energies, in contrast to the spin waves in the isotropic 
Heisenberg model. At the $\Gamma$-point, one mode has the zero excitation 
energy while another has a finite energy. 
They have the same excitation energy at the M point, but still have different
energies at the X point.

\end{abstract}


\maketitle

\section{\label{sect.1}Introduction}

The $5d$ transition-metal compounds have recently attracted much interest,
since the interplay between the spin-orbit interaction (SOI) and the electron 
correlation strongly influences their electronic structures in such systems. 
A typical example is Sr$_2$IrO$_4$, which consists of two-dimensional IrO$_2$
layers, similar to the parent compound  of cuprates La$_2$CuO$_4$.
\cite{Crawford1994}
Sr$_2$IrO$_4$ is a canted antiferromagnet below 230 K.
\cite{Crawford1994,Cao1998,Moon2006}
The electronic structure has been calculated by the LDA+U method,
\cite{Kim2008} and by the LDA combined with the dynamical mean-field theory.
\cite{Arita2012} They have found that the system is an antiferromagnetic
insulator with a finite gap in the electron-hole pair creation,
when the SOI is taken into account. The similar conclusion has been derived
by the variational Monte-Carlo calculation in the Hubbard model.
\cite{Watanabe2010}

Such spin-orbit induced antiferromagnetic insulator could be obtained 
from the localized electron picture.
In contrast to $(3d)^9$ configuration of Cu atom in La$_2$CuO$_4$, 
five $5d$ electrons are occupied per Ir atom, 
and the energy of the $e_g$ orbitals is about 2 eV higher than the energy of 
the $t_{2g}$  orbitals due to the large crystal field. Therefore, one could
regard the situation as one hole is sitting on the t$_{2g}$ orbitals.
The matrices of the orbital angular momentum operators with $L=2$ represented 
by the $t_{2g}$ states are the negative of those with $L=1$ represented by 
$|p_x\rangle$, $|p_y\rangle$, $|p_z\rangle$, if they are identified by
$|yz\rangle$, $|zx\rangle$, $|xy\rangle$, respectively, 
where $yz$, $zx$, and $xy$ designate $t_{2g}$ orbitals.
Therefore, under the SOI, the lowest-energy states of a hole are Kramers' 
doublet with the effective total angular momentum $j_{\rm eff}=1/2$:
$\frac{1}{\sqrt{3}}\left(|yz,\mp\sigma\rangle\pm i|zx,\mp\sigma\rangle
\pm|xy,\pm\sigma\rangle \right)$, where spin component $\sigma=\uparrow$
and $\downarrow$.
\cite{Kim2008,Kim2009}

The degeneracy is lifted by the inter-site interaction.
In the strong Coulomb interaction, the effective spin Hamiltonian describing 
the low-lying excitations is derived by the second-order perturbation with 
respect to the electron-transfer terms. 
Introducing the isospin operators acting on the doublet,
we obtain the Heisenberg Hamiltonian with the antiferromagnetic coupling,
consistent with the above findings.\cite{Jackeli2009,Jin2009,Kim2012}
The effect of lattice distortion has also been analyzed.\cite{Wang2011}
More importantly, it has been pointed out that the small anisotropic 
terms emerge in addition to the isotropic term, when Hund's coupling 
is taken into account 
on the two-hole states in the intermediate state of the second-order 
perturbation.\cite{Jackeli2009,Kim2012}
Such anisotropic terms are expected to modify substantially 
the excitation spectra, but have not been fully investigated yet.
The purpose of this paper is to study such effects theoretically.

We derive the effective spin Hamiltonian from 
the multi-orbital Hubbard model by taking full account of the Coulomb 
interaction in the intermediate state of the second-order perturbation. 
We obtain the exchange couplings consistent with the previous studies. 
\cite{Jackeli2009,Kim2012}
Since the anisotropic terms favor the staggered moment lying in the $ab$ plane,
we assume that the staggered moment directs to the $a$ axis.
Expanding the spin operators in terms of boson operators within the lowest 
order of $1/S$,\cite{Holstein1940} 
we introduce the Green's functions for the boson operators,
which include the so-called anomalous type.\cite{Bulut1989} 
We solve the coupled equations of motion to obtain the Green's functions.
It is found that the ``spin waves" in the isotropic Heisenberg model
are split into two modes with slightly different energy, due to the anisotropic
terms in the entire Brillouin zone. At the $\Gamma$ point, one mode has zero
excitation energy while the other has a finite energy. 
These excitation modes are to be clarified in future experiments.

This paper is organized as follows.
In Sec. II, we introduce the multi-orbital Hubbard model in the square lattice,
and derive the effective spin Hamiltonian by the second-order perturbation.
In Sec. III, we expand the spin operators in terms of boson operators, and
solve the Green's functions for boson operators. The excitation modes are
discussed. Section IV is devoted to the concluding remarks.

\section{\label{sect.2}Spin Hamiltonian for 
S\lowercase{r}$_2$I\lowercase{r}O$_4$}

\subsection{Multi-orbital Hubbard model}

The crystal structure of Sr$_2$IrO$_4$ belongs to the K$_2$NiF$_4$ type.
\cite{Crawford1994}
The oxygen octahedra surrounding an Ir atom are rotated about the 
crystallographic $c$ axis by about 11(deg).
To take account of this crystal distortion, we describe the base states 
in the local coordinate frames rotated in accordance with the rotation
of the octrahedra.\cite{Jackeli2009,Wang2011} 
Since the crystal field energy of the $e_g$ orbitals is
about 2 eV higher than that of the $t_{2g}$ orbitals, we consider only $t_{2g}$
orbitals. Electrons transfer between them
at neighboring Ir sites in the square lattice. 
Then, the multi-orbital Hubbard model is defined by
\begin{equation}
 H =  H_{\rm kin}+H_{\rm SO}+H_{\rm I},
\end{equation}
with
\begin{eqnarray}
H_{\rm kin} & = & \sum_{\left\langle i,i'\right\rangle }
\sum_{n,n'\sigma}\left(t_{in,i'n'}d_{in\sigma}^{\dagger}d_{i'n'\sigma}+ {\rm H.c.}
\right),\\
H_{\rm SO} & = & \zeta_{\rm SO}\sum_{i,n,n',\sigma,\sigma'}
d_{in\sigma}^{\dagger}({\bf L})_{nn'}
 \cdot({\bf S})_{\sigma\sigma'}d_{in'\sigma'}, \\
H_{\rm I} & = & 
 U\sum_{i,n} n_{in\uparrow}n_{in\downarrow} \nonumber \\
&+&\sum_{i,n<n'\sigma}[U' n_{in\sigma}n_{in'-\sigma}
                 + (U'-J) n_{in\sigma}n_{in'\sigma}] \nonumber\\
&+&J\sum_{i,n\neq n',\sigma} (d_{in\uparrow}^{\dagger}d_{in'\downarrow}^{\dagger}
                     d_{in\downarrow}d_{in'\uparrow}
                    +d_{in\uparrow}^{\dagger}d_{in\downarrow}^{\dagger}
                     d_{in'\downarrow}d_{in'\uparrow}), \nonumber \\
\end{eqnarray}
where $d_{in\sigma}$ denotes the annihilation operator of 
an electron with orbital $n$ ($=yz,zx,xy$) and spin $\sigma$ at the Ir site $i$.
The $H_{\rm kin}$ represents the kinetic energy with transfer integral 
$t_{in,i'n'}$, 
An electron on the $xy$ orbital could transfer to the $xy$ orbital 
in the nearest neighbor sites through the intervening O $2p$ orbitals,
while an electron on the $yz$($zx$) orbital could transfer to 
the $yz$($zx$) orbital in the nearest neighbor sites only along the $y$($x$) 
direction. 
The $H_{\rm SO}$ represents the spin-orbit interaction of $5d$ electrons
with ${\bf L}$ and ${\bf S}$ denoting the orbital and spin angular momentum 
operators.
The $H_{\rm I}$ represents the Coulomb interaction between electrons,
which satisfies $U=U'+2J$.\cite{Kanamori1963}

\subsection{Strong coupling approach}

Five electrons are occupied on $t_{2g}$ orbitals in each Ir atoms.
This state could be considered as occupying one {\emph hole}.
The matrices of the orbital angular momentum operators with $L=2$ represented by 
the $t_{2g}$ states are the minus of those with $L=1$ represented by 
$|p_x\rangle$, $|p_y\rangle$, $|p_z\rangle$, if the bases are identified by
$|yz\rangle$, $|zx\rangle$, $|xy\rangle$, respectively.
Therefore, the six-fold degenerate states are split 
into the states with the effective angular momentum $j_{\rm eff}=1/2$ and with $3/2$ 
under $H_{\rm SO}$.
The lowest-energy states are the doublet with $j_{\rm eff}=1/2$, given by
\begin{eqnarray}
 \left|+\frac{1}{2} \right\rangle &=& \frac{1}{\sqrt{3}}\left[
  |xy\uparrow \rangle + |yz\downarrow \rangle + i|zx\downarrow\rangle \right],\\
 \left|-\frac{1}{2} \right\rangle &=& \frac{1}{\sqrt{3}}\left[
  -|xy\downarrow \rangle + |yz\uparrow\rangle - i|zx\uparrow\rangle\right].
\end{eqnarray}

\begin{figure}
\includegraphics[width=8.0cm]{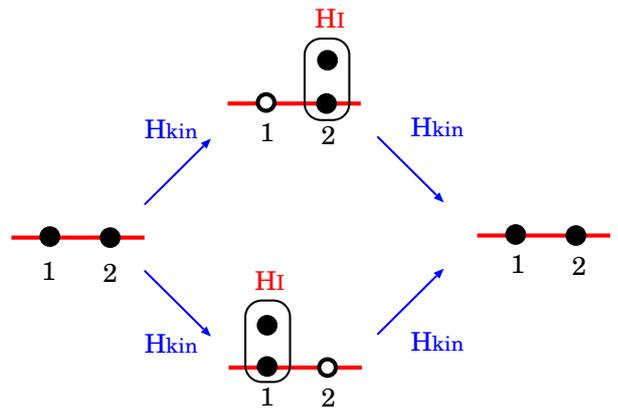}%
\caption{\label{fig.process}
(Color online) The second-order process with $H_{\rm kin}$. 
In the initial and final states, one hole sits at site 1 and another 
sits at site 2.
In the intermediate state, two holes are on the same site,
where the Coulomb interaction works.
}
\end{figure}

We start by one hole sitting at site 1 and another at site 2, and
carry out the second-order perturbation calculation, as illustrated 
in Fig.~\ref{fig.process}.
The transfer integral between $xy$ orbitals in the nearest-neighbor sites
may be generally different from those between $yz$ orbitals and between $zx$ 
orbitals, since the orbitals are defined in the local coordinate frames.
Nevertheless we assume them to be a same value, which is denoted as $t_1$,
since the difference merely gives rise to minor corrections to the values of
$J'_{z}$ and $J'_{xy}$ in Eqs. (\ref{eq.Hz}) and (\ref{eq.Hxy}). 

In the intermediate state, we take full account of the Coulomb interaction 
between two holes. We numerically evaluate the second-order energy given
in a $4\times 4$ matrix form. This matrix is expressed in terms of
the spin operators ${\bf S}$ acting on the doublet. 
It is given by 
\begin{equation}
 H(1,2)=C+H^{(0)}(1,2) + H_{z}^{(1)}(1,2) + H_{xy}^{(1)}(1,2),
\end{equation}
with
\begin{eqnarray}
 H^{(0)}(1,2) &=& J_{\rm ex}{\bf S}_1\cdot{\bf S}_2 ,\\
 H_z^{(1)}(1,2) &=& J'_z  S_1^z S_2^z, 
\label{eq.Hz}\\
 H_{xy}^{(1)}(1,2) &=& \textrm{sgn}(1,2)
J'_{xy} \left(S_1^x S_2^x - S_1^y S_2^y\right), 
\label{eq.Hxy}
\end{eqnarray}
where $\textrm{sgn}(i,j)$ gives $+1 (-1)$ when the bond 
between the sites 
$i$ and $j$ is along the $x$ ($y$) axis, and $C$ is
a constant.
Table \ref{table.1} shows the calculated coupling constants for various parameter 
sets of the Hubbard model. Both $J'_z$ and $J'_{xy}$ vanish 
without Hund's coupling $J$ because we can check 
they are proportional to $J$. Note that $J'_{z}$ is negative and 
its absolute value is nearly the same as $J'_{xy}$ within the significant 
figures. These tendencies are consistent with the previous study.
\cite{Jackeli2009,Kim2012}  

\begin{table}
\caption{\label{table.1}
Exchange couplings for various parameter sets, in units of eV. 
The transfer integral and the spin-orbit coupling are fixed at
$t_1=0.36$ and $\zeta_{\rm SO}=0.36$.}
\begin{ruledtabular}
\begin{tabular}{rrrrrr}
$U$ & $U'$   & $J$   & $J_{\rm ex}$ & $J'_{\rm z}$ & $J'_{\rm xy}$ \\
\hline
 $1.4$ & $1.4$   & $0$    & $0.165$ & $0$       & $0$      \\ 
 $1.4$ & $0.98$  & $0.21$ & $0.223$ & $-0.0055$ & $0.0055$ \\ 
\hline
 $2.2$ & $2.2$   & $0$    & $0.105$ & $0$       & $0$      \\
 $2.2$ & $1.78$  & $0.21$ & $0.124$ & $-0.0023$ & $0.0023$      \\
 $2.2$ & $1.54$  & $0.33$ & $0.144$ & $-0.0046$ & $0.0046$ \\ 
\hline
 $3.0$ & $3.0$   & $0$    & $0.077$ & $0$       & $0$      \\
 $3.0$ & $2.34$  & $0.33$ & $0.095$ & $-0.0023$ & $0.0023$ \\
 $3.0$ & $2.1$   & $0.45$ & $0.107$ & $-0.0039$ & $0.0039$ \\
\end{tabular}
\end{ruledtabular}
\end{table}

\section{Excitation Spectra} 

As a next step to the analysis of the preceding section,
we consider the following spin Hamiltonian in the square lattice: 
\begin{equation}
 H= H^{(0)} + H^{(1)},
\end{equation}
with
\begin{eqnarray}
 H^{(0)}&=&\sum_{\langle i,j\rangle} H^{(0)}(i,j), \\
 H^{(1)}&=&\sum_{\langle i,j\rangle} H_z^{(1)}(i,j) + H_{xy}^{(1)}(i,j).
\end{eqnarray}

The ground state takes the conventional antiferromagnetic spin configuration 
in the absence of the anisotropic term $H^{(1)}$. The direction of the staggered
moment is not determined. The $H_{z}^{(1)}$ makes the direction favor the 
the $xy$ plane when $J'_z < 0$.
This antiferromagnetic order breaks the rotational invariance of the isospin
space in the $ab$ plane.
We assume the staggered moment pointing to the $x$ axis.\cite{Cao1998}
It should be noted here that the antiferromagnetic order in the local coordinate
frames indicates the presence of the weak ferromagnetic moment in the global
coordinate frame. Labeling the $x$, $y$, and $z$ axes as $z'$, $x'$, and $y'$ 
axes, respectively, we express the spin operators by boson operators 
within the lowest order of $1/S$-expansion:\cite{Holstein1940}
\begin{eqnarray}
 S_i^{z'} &=& S - a_i^\dagger a_i ,  \quad 
 S_i^{x'}+iS_i^{y'} = \sqrt{2S}a_i , \label{eq.boson1}\\
 S_j^{z'} &=& -S + b_j^\dagger b_j , \quad
 S_j^{x'}+iS_j^{y'} = \sqrt{2S}b_j^\dagger ,\label{eq.boson2}
\end{eqnarray}
where $a_i$ and $b_j$ are boson annihilation operators,
and $i$ ($j$) refers to sites on the A (B) sublattice. 
Using Eqs.~(\ref{eq.boson1}) and (\ref{eq.boson2}),
$H^{(0)}$ and $H^{(1)}$ may be expressed as
\begin{eqnarray}
 H^{(0)} &=& J_{\rm ex}S\sum_{\langle i,j \rangle}( a_i^\dagger a_i + b^\dagger_{j} b_{j}
           +  a_i b_{j} + a_i^\dagger b_{j}^\dagger), 
\label{eq.h0} \\
 H^{(1)} &=& J'_z S\frac{1}{2}\sum_{\langle i,j \rangle}
            (a_i - a_i^\dagger)(b_j - b_j^\dagger) \nonumber\\
 &+&J'_{xy}S \sum_{\langle i,j \rangle}\textrm{sgn}(i,j)
(a_i^{\dagger}a_i + b_j^{\dagger}b_j) \nonumber \\
& -&J'_{xy}S\frac{1}{2} \sum_{\langle i,j \rangle}
            \textrm{sgn}(i,j)(a_i +a_i^{\dagger})(b_j^{\dagger}+b_j).
\label{eq.h1}
\end{eqnarray} 
where the unimportant constant term is neglected. 
The second term in Eq.~(\ref{eq.h1}) is canceled out by the factor $\pm$.
Then we introduce the Fourier transforms of the boson operators in the 
magnetic Brillouin zone, 
\begin{eqnarray}
 a({\bf k}) &=& \sqrt{\frac{2}{N}}\sum_{i}a_{i}
                             \exp(-i{\bf k}\cdot{\bf r}_i) , \\
 b({\bf k}) &=& \sqrt{\frac{2}{N}}\sum_{j}b_{j}
                             \exp(-i{\bf k}\cdot{\bf r}_j) ,
\label{eq.Fouriers2}
\end{eqnarray}
where $N$ is the number of sites, and $i$ ($j$) runs over A (B) sublattice. 
We obtain
\begin{eqnarray}
 H^{(0)} &=& J_{\rm ex}Sz\sum_{\bf k} a^{\dagger}({\bf k})a({\bf k}) 
  + b^{\dagger}({\bf k})b({\bf k})
 \nonumber \\
& +& \gamma({\bf k})[a^{\dagger}({\bf k})b^{\dagger}({\bf -k})
                   +a({\bf k})b({\bf -k})],
\\
 H_{z}^{(1)} &=& J'_{z}(2S)\sum_{\bf k} \gamma({\bf k})
    [a({\bf k})-a^{\dagger}({\bf -k})][b({\bf -k}) - b^{\dagger}({\bf k})], \nonumber \\
\\
 H_{xy}^{(1)} &=& -J'_{xy}(2S)\sum_{\bf k} \eta({\bf k})
 [a({\bf k})+a^{\dagger}({\bf -k})][b({\bf -k})+b^{\dagger}({\bf k})], 
\nonumber \\
\end{eqnarray}
where
\begin{eqnarray}
 \gamma({\bf k}) &=& \frac{1}{2}(\cos k_x + \cos k_y), \\
 \eta({\bf k}) &=& \frac{1}{2}(\cos k_x - \cos k_y).
\end{eqnarray}
Here $z$ is the number of nearest neighbors, {\it i.e.}, $z=4$.

To find out the excitation modes, we introduce the Green's functions,
\begin{eqnarray}
 G_{aa}({\bf k},t) &=& -i\langle T[a({\bf k},t)a^{\dagger}({\bf k},0)]\rangle, \\
 F_{ba}({\bf k},t) &=& -i\langle T[b^{\dagger}({\bf -k},t)a^{\dagger}({\bf k},0)]\rangle, \\
 G_{ba}({\bf k},t) &=& -i\langle T[b({\bf k},t)a^{\dagger}({\bf k},0)]\rangle, \\
 F_{aa}({\bf k},t) &=& -i\langle T[a^{\dagger}({\bf -k},t)a^{\dagger}({\bf k},0)]\rangle,
\end{eqnarray}
where $T$ is the time ordering operators, and $\langle X \rangle$ denotes the ground-state
average of operator $X$. The $F_{ba}({\bf k},t)$ and $F_{aa}({\bf k},t)$ belong to the 
so called anomalous type. Defining their Fourier transforms by
$G_{aa}({\bf k},\omega) = \int G_{aa}({\bf k},t){\rm e}^{i\omega t}{\rm d}t$
and so on, we derive the equation of motion for these functions,
\begin{eqnarray}
 &&\left( \begin{array}{cccc}
  \omega -1 & -A({\bf k}) & B({\bf k}) & 0 \\
  -A({\bf k})  & -(\omega+1) & 0 & B({\bf k}) \\
  B({\bf k}) & 0 & \omega-1 & -A({\bf k}) \\
  0 & B({\bf k}) & -A({\bf k}) & -(\omega+1) 
  \end{array} \right)
\nonumber \\
&\times&   \left( \begin{array}{c}
         G_{aa}({\bf k},\omega) \\
         F_{ba}({\bf k},\omega) \\
         G_{ba}({\bf k},\omega) \\
         F_{aa}({\bf k},\omega)
        \end{array} \right)
= \left( \begin{array}{c}
            1 \\
            0 \\
            0 \\
            0 
           \end{array} \right) , \label{eq.matrix}
\end{eqnarray}
where 
\begin{eqnarray}
   A({\bf k}) &=& (1+g_z)\gamma({\bf k}) -g_{xy}\eta({\bf k}) , \\
   B({\bf k}) &=& g_z\gamma({\bf k}) + g_{xy}\eta({\bf k}) , \\
   g_z &=& J'_{z}/(2J_{\rm ex}), \quad g_{xy}=J'_{xy}/(2J_{\rm ex}). 
\end{eqnarray}
Here the energy is measured in units of $J_{\rm ex}Sz$.
Hence we finally obtain,
\begin{equation}
\left( \begin{array}{c}
 G_{aa}({\bf k},\omega) \\
 F_{ba}({\bf k},\omega) \\
 G_{ba}({\bf k},\omega) \\
 F_{aa}({\bf k},\omega) \\
\end{array} \right)=\frac{1}{D({\bf k},\omega)}
\left( \begin{array}{c}
g_{aa}({\bf k},\omega) \\
f_{ba}({\bf k},\omega) \\
g_{ba}({\bf k},\omega) \\
f_{aa}({\bf k},\omega) \\
\end{array} \right), 
\end{equation}
where
\begin{eqnarray}
 D({\bf k},\omega) &=& \omega^4 -2[1+B({\bf k})^2 - A({\bf k})^2]\omega^2
 +1-2A({\bf k})^2 \nonumber \\
&-&2B({\bf k})^2+[A({\bf k})^2-B({\bf k})^2]^2, 
\label{eq.determ} \\
 g_{aa}({\bf k},\omega) &=& (\omega-1)(\omega+1)^2 - B({\bf k})^2(\omega-1)
 \nonumber \\
&+&A({\bf k})^2(\omega+1) ,\\
 f_{ba}({\bf k},\omega) &=& -A({\bf k})[(\omega^2-1)-B({\bf k})^2
        +A({\bf k})^2], \\
 g_{ba}({\bf k},\omega) &=& B({\bf k})
      [B({\bf k})^2-(\omega+1)^2-A({\bf k})^2], \\
 f_{aa}({\bf k},\omega) &=& 2A({\bf k})B({\bf k}).
\label{eq.faa}
\end{eqnarray}
In the absence of the anisotropic terms, we have 
$A({\bf k})=\gamma({\bf k})$ and $B({\bf k})=0$. Inserting these relations
into Eqs.~(\ref{eq.determ})-(\ref{eq.faa}), we have
\begin{eqnarray}
 G_{aa}({\bf k},\omega)&=&\frac{\omega+1}{\omega^2-(1-\gamma({\bf k})^2)}, \\
 F_{ba}({\bf k},\omega)&=&-\frac{\gamma({\bf k})}{\omega^2-(1-\gamma({\bf k})^2)},\\
 G_{ba}({\bf k},\omega)&=& F_{aa}({\bf k},\omega) = 0.
\end{eqnarray}
These forms are well-known for the isotropic Heisenberg model.\cite{Bulut1989}

In the presence of the anisotropic terms, Eq.~(\ref{eq.determ}) is 
rewritten as
\begin{equation}
D(\textbf{k},\omega) =[ \omega^2 - E_-^2(\textbf{k})]
[\omega^2 - E_+^2(\textbf{k})],
\end{equation}
with
\begin{equation}
 E_{\pm}(\textbf{k}) = \sqrt{ [1 \pm |B(\textbf{k})|]^2 - A^2(\textbf{k}) }.
\label{eq.disp.1}
\end{equation}
This indicates that poles exist at $\omega=E_{\pm}({\bf k})$ 
in the domain of $\omega> 0$.
To clarify the behavior of the poles, we express the Green's function as
\begin{equation}
G_{aa}(\textbf{k},\omega)=
\frac{1}{2} \left[
  \frac{\omega + 1 +|B(\textbf{k})|}
                        {\omega^2-E_+^2(\textbf{k})}
+ \frac{\omega + 1 -|B(\textbf{k})|}
                        {\omega^2-E_-^2(\textbf{k})}
\right].
\end{equation}
This form indicates that two poles have nearly equal weights in the 
domain of
$\omega>0$ in the case of weak anisotropic terms.
At the $\Gamma$-point, $A({\bf k})=1+g_z$ and $B({\bf k})=g_z$, and hence 
$D(0,\omega)$ $=$ $\omega^2(\omega^2+4g_z)$. Therefore one mode has
zero excitation energy while the other has a finite energy $2\sqrt{-g_z}$. 
The former may correspond to the Goldstone mode due to
breaking the rotational invariance of the isospin in the $ab$ plane.
At the X-point, ${\bf k}=(\pi,0)$,  $A({\bf k})=$ $-B({\bf k})$ $=g_{xy}$, 
and hence the two modes have excitation energies 
$\omega=\sqrt{1\pm 2g_{xy}}$.
At the M-point, since $A({\bf k})=$ $B({\bf k})=0$, the two modes have 
the same excitation energy, $\omega=1$.

\begin{figure}
\includegraphics[width=8.0cm]{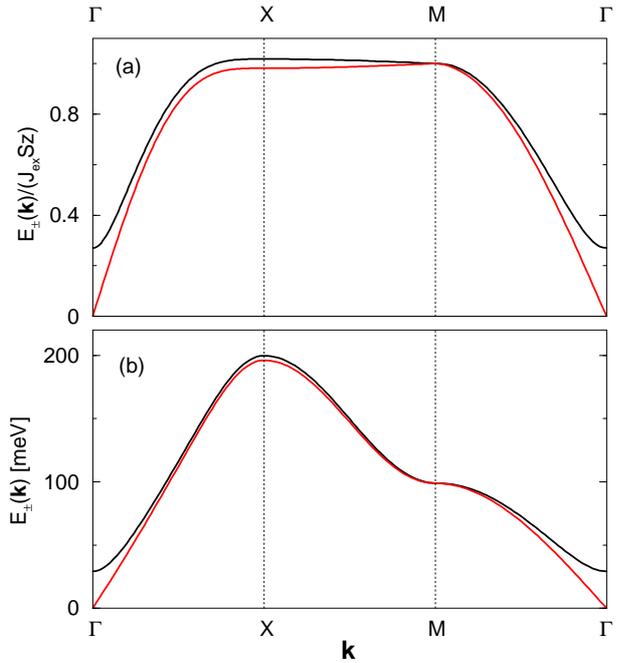}%
\caption{\label{fig.dispersion}
(Color online) Excitation energies of two modes $E_{\pm}(\textbf{k})$
as a function of ${\bf k}$ along the symmetry lines. 
(a) $E_{\pm}(\textbf{k})$ evaluated from Eq. (\ref{eq.disp.1}).
The parameters are evaluated from the Hubbard model with
$t_1=0.36$ eV, $\zeta_{\rm SO}=0.36$ eV, $U=3.0$ eV,
$U'=2.1$ eV, and $J=0.45$ eV with 
$g_{z}=-0.018$ and $g_{xy}=0.018$ 
in units of $J_{\rm ex}Sz$.
(b) $E_{\pm}(\textbf{k})$ evaluated from Eq. (\ref{eq.disp.2}) including
$J_{\textrm{ex}}'/J_{\textrm{ex}}=-1/3$ and 
$J_{\textrm{ex}}''/J_{\textrm{ex}}=1/4$ with $J_{\textrm{ex}}=59$ meV.
The parameters are evaluated from the Hubbard model with
$t_1=0.3$ eV, $\zeta_{\rm SO}=0.4$ eV, $U=3.5$ eV,
$U'=2.6$ eV, and $J=0.45$ eV with 
$g_{z}=-0.015$ and $g_{xy}=0.015$ 
in units of $J_{\rm ex}Sz$.
}
\end{figure}

Figure \ref{fig.dispersion} shows the dispersion relation along the symmetry 
lines of ${\bf k}$ for $g_{z}=-0.018$ and $g_{xy}=0.018$.
The parameters correspond to the Hubbard model with
$t_1=0.36$ eV, $\zeta_{\rm SO}=0.36$ eV, $U=3.0$ eV,
$U'=2.1$ eV, and $J=0.45$ eV (last row in Table \ref{table.1}).
Although $J'_{z}$ and $J'_{xy}$ are two orders of magnitude smaller than 
$J_{\rm ex}$, they substantially modify the dispersion relation.
Note that $J_{\rm ex}Sz(\equiv 2J_{\rm ex})=0.214$ eV, which is comparable 
to the excitation energy at the X-point observed in resonant inelastic
x-ray scattering (RIXS) at the $L$-edge of Ir.\cite{J.Kim2012}

Finally, let us consider what happens 
when the exchange interactions between the second and third
neighbor sites, denoted as $J_{\textrm{ex}}'$ and
$J_{\textrm{ex}}''$ respectively, are introduced 
in addition to $J_{\textrm{ex}}$.
It is known that a phenomenological isotropic model
constructed by $J_{\textrm{ex}}$, $J_{\textrm{ex}}'$, and 
$J_{\textrm{ex}}''$
couplings gives much better dispersion curve.\cite{J.Kim2012}
We expect the inclusion of $J_{\textrm{ex}}'$ and $J_{\textrm{ex}}''$ terms
improves the dispersion curve
in comparison with the experimental one since the contributions from the anisotropic terms are not so significant in the wide range of the 
Brillouin zone as shown in Fig. \ref{fig.dispersion} (a).
Then, our concern is whether or not the gap between the two modes 
around the $\Gamma$ point remains finite quantitatively.
We see the answer is in the affirmative as follows.

In the presence of $J_{\textrm{ex}}'$ and $J_{\textrm{ex}}''$ terms,
the coefficient matrix appeared in Eq. (\ref{eq.matrix})
is modified as
\begin{widetext}
\begin{equation}
 \left( \begin{array}{cccc}
  \omega -1+\xi(\textbf{k}) & -A({\bf k}) & B({\bf k}) & 0 \\
  -A({\bf k})  & -(\omega+1-\xi(\textbf{k})) & 0 & B({\bf k}) \\
  B({\bf k}) & 0 & \omega-1+\xi(\textbf{k}) & -A({\bf k}) \\
  0 & B({\bf k}) & -A({\bf k}) & -(\omega+1-\xi(\textbf{k})) 
  \end{array} \right),
\end{equation}
\end{widetext}
where
\begin{eqnarray}
\xi(\textbf{k})&=& \frac{J_{\textrm{ex}}'}{J_{\textrm{ex}}}
( 1-\gamma'(\textbf{k}))
+ \frac{J_{\textrm{ex}}''}{J_{\textrm{ex}}}( 1-\gamma''(\textbf{k})), \\
\gamma'(\textbf{k}) &=& \cos k_x \cos k_y, \\
\gamma''(\textbf{k}) &=& \frac{1}{2} [ \cos(2k_x) + \cos(2 k_y)].
\end{eqnarray}
Then, the excitation energies for two modes become
\begin{equation}
 E_{\pm}(\textbf{k}) = 
\sqrt{ [1 - \xi(\textbf{k})
\pm |B(\textbf{k})|]^2 - A^2(\textbf{k}) }. \label{eq.disp.2}
\end{equation}
Since the extra term $\xi(\textbf{k})$ goes to zero when
$|\textbf{k}| \rightarrow 0$, the existence of the gap
of the two modes at the $\Gamma$ point is robust.

The experimental dispersion curve can be reproduced
well by setting $J_{\textrm{ex}}=60$ meV, 
$J_{\textrm{ex}}'=-J_{\textbf{ex}}/3$ and 
$J_{\textrm{ex}}''=J_{\textbf{ex}}/4$ in the phenomenological model
without the anisotropic terms $J_z '$ and $J_{xy}'$.\cite{J.Kim2012}
With the parameter set of $U=3.5$ eV, $U'=2.6$ eV, $J=0.45$ eV, 
$t_1=0.3$ eV, and $\zeta_{\textrm{SO}}=0.4$ eV, 
we obtain $J_{\textrm{ex}}\simeq 60$ meV, $J_{xy}'=-J_z'=1.8$ meV.
Together with the relations $J_{\textrm{ex}}'=-J_{\textbf{ex}}/3$ and 
$J_{\textrm{ex}}''=J_{\textbf{ex}}/4$, the $E_{\pm}(\textbf{K})$
can be numerically evaluated. 
As shown in Fig. \ref{fig.dispersion} (b), the inclusion of 
$J_{\textrm{ex}}'$ and $J_{\textrm{ex}}''$ 
improves the dispersion curve as expected. On the other hand, the 
splitting of the two modes, which is the most prominent around
the $\Gamma$ point, remains nearly intact and the magnitude of the
gap keeps nearly the same order of magnitude as that obtained for
the parameter set evaluated in the absence of 
$J_{\textrm{ex}}'$ and $J_{\textrm{ex}}''$.

\section{Concluding remarks}

We have studied the low-lying excitations in Sr$_2$IrO$_4$ on
the localized electron picture. Having introduced the isospin operators
acting on Kramers' doublet, we have derived the effective spin Hamiltonian 
from the multi-orbital Hubbard model by the second-order perturbation with 
the electron transfer.
This approach may be justified in the strong Coulomb interaction.
It consists of the isotropic Heisenberg term and small anisotropic terms.
Expanding the spin operators in terms of boson operators, we have introduced
the Green's functions for the boson operators, and have solved the coupled
equations of motion for those functions.
The excitation spectra have been obtained from the Green's functions.
It is found that two modes emerge with slightly different energies due to
the anisotropic terms, in contrast to the spin waves in the isotropic 
Heisenberg model.

The existence of the anisotropic terms, which is the hallmark of 
the interplay between the SOI and the Coulomb interaction,
has not been corroborated by experiments.\cite{J.Kim2012,Fujiyama2012} 
The magnetic excitations are usually detected by the inelastic neutron 
scattering, but it may be hard for the present system due to the
strong absorption of neutron by Ir atom.
Recently, the spin and orbital excitations have been observed and analyzed
by the Ir $L$-edge RIXS. No indication of the splitting is unfortunately
found in the spectral shape.\cite{J.Kim2012,Ament2011}
Since the energy difference is around $60$ meV at most,
it may be difficult to distinguish the two modes by the RIXS experiments.
However, the observation with finer instrumental energy resolution 
of 36 meV may be within the reach at the $L_{2,3}$ edges of 
Ir.\cite{MorettiSala2013}

Finally we comment on the validity of the strong coupling approach. 
The excitation energy at the M point is the same as at the X point
within the nearest-neighbor coupling in the Heisenberg model. 
Since
the energy at the M point is found nearly the half of that at 
the X point in the RIXS, we need to include the next-nearest-neighbor
coupling as large as one third of the nearest neighbor coupling to 
account for such difference.\cite{J.Kim2012}
In addition, the Mott-Hubbard gap is estimated as $\sim 0.4$ eV from the
optical absorption spectra,\cite{Kim2008,Moon2009}
which value is comparable to the magnetic excitation energy $2J\sim 0.2$ eV.
These indicate that the strong coupling approach may not
work well.
It may be interesting to study the elementary excitations from the viewpoint
of the itinerant electron picture workable in the weak and 
intermediate couplings.

\begin{acknowledgments}
We are grateful to M. Yokoyama for fruitful discussions.
This work was partially supported by a Grant-in-Aid for Scientific Research
from the Ministry of Education, Culture, Sports, Science and Technology
of the Japanese Government.
\end{acknowledgments}

\bibliographystyle{apsrev} \bibliographystyle{apsrev}
\bibliography{paper}

\begin{thebibliography}{19}
\expandafter\ifx\csname natexlab\endcsname\relax\def\natexlab#1{#1}\fi
\expandafter\ifx\csname bibnamefont\endcsname\relax
  \def\bibnamefont#1{#1}\fi
\expandafter\ifx\csname bibfnamefont\endcsname\relax
  \def\bibfnamefont#1{#1}\fi
\expandafter\ifx\csname citenamefont\endcsname\relax
  \def\citenamefont#1{#1}\fi
\expandafter\ifx\csname url\endcsname\relax
  \def\url#1{\texttt{#1}}\fi
\expandafter\ifx\csname urlprefix\endcsname\relax\def\urlprefix{URL }\fi
\providecommand{\bibinfo}[2]{#2}
\providecommand{\eprint}[2][]{\url{#2}}

\bibitem[{\citenamefont{Crawford et~al.}(1994)\citenamefont{Crawford,
  Subramanian, Harlow, Fernandez-Baca, Wang, and Johnston}}]{Crawford1994}
\bibinfo{author}{\bibfnamefont{M.~K.} \bibnamefont{Crawford}},
  \bibinfo{author}{\bibfnamefont{M.~A.} \bibnamefont{Subramanian}},
  \bibinfo{author}{\bibfnamefont{R.~L.} \bibnamefont{Harlow}},
  \bibinfo{author}{\bibfnamefont{J.~A.} \bibnamefont{Fernandez-Baca}},
  \bibinfo{author}{\bibfnamefont{Z.~R.} \bibnamefont{Wang}}, \bibnamefont{and}
  \bibinfo{author}{\bibfnamefont{D.~C.} \bibnamefont{Johnston}},
  \bibinfo{journal}{Phys.\ Rev.\ B} \textbf{\bibinfo{volume}{49}},
  \bibinfo{pages}{9198} (\bibinfo{year}{1994}).

\bibitem[{\citenamefont{Cao et~al.}(1998)\citenamefont{Cao, Bolivar, McCall,
  Crow, and Guertin}}]{Cao1998}
\bibinfo{author}{\bibfnamefont{G.}~\bibnamefont{Cao}},
  \bibinfo{author}{\bibfnamefont{J.}~\bibnamefont{Bolivar}},
  \bibinfo{author}{\bibfnamefont{S.}~\bibnamefont{McCall}},
  \bibinfo{author}{\bibfnamefont{J.~E.} \bibnamefont{Crow}}, \bibnamefont{and}
  \bibinfo{author}{\bibfnamefont{R.~P.} \bibnamefont{Guertin}},
  \bibinfo{journal}{Phys.\ Rev.\ B} \textbf{\bibinfo{volume}{57}},
  \bibinfo{pages}{R11039} (\bibinfo{year}{1998}).

\bibitem[{\citenamefont{Moon et~al.}(2006)\citenamefont{Moon, Kim, Kim, Lee,
  Kim, Park, Kim, Oh, Nakatsuji, Maeno et~al.}}]{Moon2006}
\bibinfo{author}{\bibfnamefont{S.~J.} \bibnamefont{Moon}},
  \bibinfo{author}{\bibfnamefont{M.~W.} \bibnamefont{Kim}},
  \bibinfo{author}{\bibfnamefont{K.~W.} \bibnamefont{Kim}},
  \bibinfo{author}{\bibfnamefont{Y.~S.} \bibnamefont{Lee}},
  \bibinfo{author}{\bibfnamefont{J.-Y.} \bibnamefont{Kim}},
  \bibinfo{author}{\bibfnamefont{J.-H.} \bibnamefont{Park}},
  \bibinfo{author}{\bibfnamefont{B.~J.} \bibnamefont{Kim}},
  \bibinfo{author}{\bibfnamefont{S.-J.} \bibnamefont{Oh}},
  \bibinfo{author}{\bibfnamefont{S.}~\bibnamefont{Nakatsuji}},
  \bibinfo{author}{\bibfnamefont{Y.}~\bibnamefont{Maeno}},
  \bibnamefont{et~al.}, \bibinfo{journal}{Phys.\ Rev.\ B}
  \textbf{\bibinfo{volume}{74}}, \bibinfo{pages}{113104}
  (\bibinfo{year}{2006}).

\bibitem[{\citenamefont{Kim et~al.}(2008)\citenamefont{Kim, Jin, Moon, Kim,
  Park, Leem, Yu, Noh, Kim, Oh et~al.}}]{Kim2008}
\bibinfo{author}{\bibfnamefont{B.~J.} \bibnamefont{Kim}},
  \bibinfo{author}{\bibfnamefont{H.}~\bibnamefont{Jin}},
  \bibinfo{author}{\bibfnamefont{S.~J.} \bibnamefont{Moon}},
  \bibinfo{author}{\bibfnamefont{J.-Y.} \bibnamefont{Kim}},
  \bibinfo{author}{\bibfnamefont{B.-G.} \bibnamefont{Park}},
  \bibinfo{author}{\bibfnamefont{C.~S.} \bibnamefont{Leem}},
  \bibinfo{author}{\bibfnamefont{J.}~\bibnamefont{Yu}},
  \bibinfo{author}{\bibfnamefont{T.~W.} \bibnamefont{Noh}},
  \bibinfo{author}{\bibfnamefont{C.}~\bibnamefont{Kim}},
  \bibinfo{author}{\bibfnamefont{S.-J.} \bibnamefont{Oh}},
  \bibnamefont{et~al.}, \bibinfo{journal}{Phys.\ Rev.\ Lett.}
  \textbf{\bibinfo{volume}{101}}, \bibinfo{pages}{076402}
  (\bibinfo{year}{2008}).

\bibitem[{\citenamefont{Arita et~al.}(2012)\citenamefont{Arita, Kune\v{s},
  Kozhevnikov, Eguiluz, and Imada}}]{Arita2012}
\bibinfo{author}{\bibfnamefont{R.}~\bibnamefont{Arita}},
  \bibinfo{author}{\bibfnamefont{J.}~\bibnamefont{Kune\v{s}}},
  \bibinfo{author}{\bibfnamefont{A.~V.} \bibnamefont{Kozhevnikov}},
  \bibinfo{author}{\bibfnamefont{A.~G.} \bibnamefont{Eguiluz}},
  \bibnamefont{and} \bibinfo{author}{\bibfnamefont{M.}~\bibnamefont{Imada}},
  \bibinfo{journal}{Phys.\ Rev.\ Lett.} \textbf{\bibinfo{volume}{108}},
  \bibinfo{pages}{086403} (\bibinfo{year}{2012}).

\bibitem[{\citenamefont{Watanabe et~al.}(2010)\citenamefont{Watanabe,
  Shirakawa, and Yunoki}}]{Watanabe2010}
\bibinfo{author}{\bibfnamefont{H.}~\bibnamefont{Watanabe}},
  \bibinfo{author}{\bibfnamefont{T.}~\bibnamefont{Shirakawa}},
  \bibnamefont{and} \bibinfo{author}{\bibfnamefont{S.}~\bibnamefont{Yunoki}},
  \bibinfo{journal}{Phys.\ Rev.\ Lett.} \textbf{\bibinfo{volume}{105}},
  \bibinfo{pages}{216410} (\bibinfo{year}{2010}).

\bibitem[{\citenamefont{Kim et~al.}(2009)\citenamefont{Kim, Ohsumi, Komesu,
  Sakai, Morita, Takagi, and Arima}}]{Kim2009}
\bibinfo{author}{\bibfnamefont{B.~J.} \bibnamefont{Kim}},
  \bibinfo{author}{\bibfnamefont{H.}~\bibnamefont{Ohsumi}},
  \bibinfo{author}{\bibfnamefont{T.}~\bibnamefont{Komesu}},
  \bibinfo{author}{\bibfnamefont{S.}~\bibnamefont{Sakai}},
  \bibinfo{author}{\bibfnamefont{T.}~\bibnamefont{Morita}},
  \bibinfo{author}{\bibfnamefont{H.}~\bibnamefont{Takagi}}, \bibnamefont{and}
  \bibinfo{author}{\bibfnamefont{T.}~\bibnamefont{Arima}},
  \bibinfo{journal}{Science} \textbf{\bibinfo{volume}{323}},
  \bibinfo{pages}{1329} (\bibinfo{year}{2009}).

\bibitem[{\citenamefont{Jackeli and Khaliullin}(2009)}]{Jackeli2009}
\bibinfo{author}{\bibfnamefont{G.}~\bibnamefont{Jackeli}} \bibnamefont{and}
  \bibinfo{author}{\bibfnamefont{G.}~\bibnamefont{Khaliullin}},
  \bibinfo{journal}{Phys.\ Rev.\ Lett.} \textbf{\bibinfo{volume}{102}},
  \bibinfo{pages}{017205} (\bibinfo{year}{2009}).

\bibitem[{\citenamefont{Jin et~al.}(2009)\citenamefont{Jin, Jeong, Ozaki, and
  Yu}}]{Jin2009}
\bibinfo{author}{\bibfnamefont{H.}~\bibnamefont{Jin}},
  \bibinfo{author}{\bibfnamefont{H.}~\bibnamefont{Jeong}},
  \bibinfo{author}{\bibfnamefont{T.}~\bibnamefont{Ozaki}}, \bibnamefont{and}
  \bibinfo{author}{\bibfnamefont{J.}~\bibnamefont{Yu}},
  \bibinfo{journal}{Phys.\ Rev.\ B} \textbf{\bibinfo{volume}{80}},
  \bibinfo{pages}{075112} (\bibinfo{year}{2009}).

\bibitem[{\citenamefont{Kim et~al.}(2012{\natexlab{a}})\citenamefont{Kim,
  Khaliullin, and Min}}]{Kim2012}
\bibinfo{author}{\bibfnamefont{B.~H.} \bibnamefont{Kim}},
  \bibinfo{author}{\bibfnamefont{G.}~\bibnamefont{Khaliullin}},
  \bibnamefont{and} \bibinfo{author}{\bibfnamefont{B.~I.} \bibnamefont{Min}},
  \bibinfo{journal}{Phys.\ Rev.\ Lett.} \textbf{\bibinfo{volume}{109}},
  \bibinfo{pages}{167205} (\bibinfo{year}{2012}{\natexlab{a}}).

\bibitem[{\citenamefont{Wang and Senthil}(2011)}]{Wang2011}
\bibinfo{author}{\bibfnamefont{F.}~\bibnamefont{Wang}} \bibnamefont{and}
  \bibinfo{author}{\bibfnamefont{T.}~\bibnamefont{Senthil}},
  \bibinfo{journal}{Phys.\ Rev.\ Lett.} \textbf{\bibinfo{volume}{106}},
  \bibinfo{pages}{136402} (\bibinfo{year}{2011}).

\bibitem[{\citenamefont{Holstein and Primakoff}(1940)}]{Holstein1940}
\bibinfo{author}{\bibfnamefont{T.}~\bibnamefont{Holstein}} \bibnamefont{and}
  \bibinfo{author}{\bibfnamefont{H.}~\bibnamefont{Primakoff}},
  \bibinfo{journal}{Phys.\ Rev.} \textbf{\bibinfo{volume}{58}},
  \bibinfo{pages}{1098} (\bibinfo{year}{1940}).

\bibitem[{\citenamefont{Bulut et~al.}(1989)\citenamefont{Bulut, Hone,
  Scalapino, and Loh}}]{Bulut1989}
\bibinfo{author}{\bibfnamefont{N.}~\bibnamefont{Bulut}},
  \bibinfo{author}{\bibfnamefont{D.}~\bibnamefont{Hone}},
  \bibinfo{author}{\bibfnamefont{D.~J.} \bibnamefont{Scalapino}},
  \bibnamefont{and} \bibinfo{author}{\bibfnamefont{E.~Y.} \bibnamefont{Loh}},
  \bibinfo{journal}{Phys.\ Rev.\ Lett.} \textbf{\bibinfo{volume}{62}},
  \bibinfo{pages}{2192} (\bibinfo{year}{1989}).

\bibitem[{\citenamefont{Kanamori}(1963)}]{Kanamori1963}
\bibinfo{author}{\bibfnamefont{J.}~\bibnamefont{Kanamori}},
  \bibinfo{journal}{Prog.\ Theor.\ Phys.} \textbf{\bibinfo{volume}{30}},
  \bibinfo{pages}{275} (\bibinfo{year}{1963}).

\bibitem[{\citenamefont{Kim et~al.}(2012{\natexlab{b}})\citenamefont{Kim, Casa,
  Upton, Gog, Kim, Mitchell, van Veenendaal, Daghofer, van~den Brink,
  Khaliullin et~al.}}]{J.Kim2012}
\bibinfo{author}{\bibfnamefont{J.}~\bibnamefont{Kim}},
  \bibinfo{author}{\bibfnamefont{D.}~\bibnamefont{Casa}},
  \bibinfo{author}{\bibfnamefont{M.~H.} \bibnamefont{Upton}},
  \bibinfo{author}{\bibfnamefont{T.}~\bibnamefont{Gog}},
  \bibinfo{author}{\bibfnamefont{Y.-J.} \bibnamefont{Kim}},
  \bibinfo{author}{\bibfnamefont{J.~F.} \bibnamefont{Mitchell}},
  \bibinfo{author}{\bibfnamefont{M.}~\bibnamefont{van Veenendaal}},
  \bibinfo{author}{\bibfnamefont{M.}~\bibnamefont{Daghofer}},
  \bibinfo{author}{\bibfnamefont{J.}~\bibnamefont{van~den Brink}},
  \bibinfo{author}{\bibfnamefont{G.}~\bibnamefont{Khaliullin}},
  \bibnamefont{et~al.}, \bibinfo{journal}{Phys.\ Rev.\ Lett.}
  \textbf{\bibinfo{volume}{108}}, \bibinfo{pages}{177003}
  (\bibinfo{year}{2012}{\natexlab{b}}).

\bibitem[{\citenamefont{Fujiyama et~al.}(2012)\citenamefont{Fujiyama, Ohsumi,
  Komesu, Matsuno, Kim, Takata, Arima, and Takagi}}]{Fujiyama2012}
\bibinfo{author}{\bibfnamefont{S.}~\bibnamefont{Fujiyama}},
  \bibinfo{author}{\bibfnamefont{H.}~\bibnamefont{Ohsumi}},
  \bibinfo{author}{\bibfnamefont{T.}~\bibnamefont{Komesu}},
  \bibinfo{author}{\bibfnamefont{J.}~\bibnamefont{Matsuno}},
  \bibinfo{author}{\bibfnamefont{B.~J.} \bibnamefont{Kim}},
  \bibinfo{author}{\bibfnamefont{M.}~\bibnamefont{Takata}},
  \bibinfo{author}{\bibfnamefont{T.}~\bibnamefont{Arima}}, \bibnamefont{and}
  \bibinfo{author}{\bibfnamefont{H.}~\bibnamefont{Takagi}},
  \bibinfo{journal}{Phys.\ Rev.\ Lett.} \textbf{\bibinfo{volume}{108}},
  \bibinfo{pages}{247212} (\bibinfo{year}{2012}).

\bibitem[{\citenamefont{Ament et~al.}(2011)\citenamefont{Ament, Khaliullin, and
  van~den Brink}}]{Ament2011}
\bibinfo{author}{\bibfnamefont{L.~J.~P.} \bibnamefont{Ament}},
  \bibinfo{author}{\bibfnamefont{G.}~\bibnamefont{Khaliullin}},
  \bibnamefont{and} \bibinfo{author}{\bibfnamefont{J.}~\bibnamefont{van~den
  Brink}}, \bibinfo{journal}{Phys.\ Rev.\ B} \textbf{\bibinfo{volume}{84}},
  \bibinfo{pages}{020403} (\bibinfo{year}{2011}).

\bibitem[{\citenamefont{Sala et~al.}(2013)\citenamefont{Sala, Henriquet,
  Simonelli, Verbeni, and Monaco}}]{MorettiSala2013}
\bibinfo{author}{\bibfnamefont{M.~M.} \bibnamefont{Sala}},
  \bibinfo{author}{\bibfnamefont{C.}~\bibnamefont{Henriquet}},
  \bibinfo{author}{\bibfnamefont{L.}~\bibnamefont{Simonelli}},
  \bibinfo{author}{\bibfnamefont{R.}~\bibnamefont{Verbeni}}, \bibnamefont{and}
  \bibinfo{author}{\bibfnamefont{G.}~\bibnamefont{Monaco}},
  \bibinfo{journal}{J. Electron Spectrosc. Relat. Phenom.}
  \textbf{\bibinfo{volume}{188}}, \bibinfo{pages}{150} (\bibinfo{year}{2013}).

\bibitem[{\citenamefont{Moon et~al.}(2009)\citenamefont{Moon, Jin, Choi, Lee,
  Seo, Yu, Cao, Noh, and Lee}}]{Moon2009}
\bibinfo{author}{\bibfnamefont{S.~J.} \bibnamefont{Moon}},
  \bibinfo{author}{\bibfnamefont{H.}~\bibnamefont{Jin}},
  \bibinfo{author}{\bibfnamefont{W.~S.} \bibnamefont{Choi}},
  \bibinfo{author}{\bibfnamefont{J.~S.} \bibnamefont{Lee}},
  \bibinfo{author}{\bibfnamefont{S.~S.~A.} \bibnamefont{Seo}},
  \bibinfo{author}{\bibfnamefont{J.}~\bibnamefont{Yu}},
  \bibinfo{author}{\bibfnamefont{G.}~\bibnamefont{Cao}},
  \bibinfo{author}{\bibfnamefont{T.~W.} \bibnamefont{Noh}}, \bibnamefont{and}
  \bibinfo{author}{\bibfnamefont{Y.~S.} \bibnamefont{Lee}},
  \bibinfo{journal}{Phys.\ Rev.\ B} \textbf{\bibinfo{volume}{80}},
  \bibinfo{pages}{195110} (\bibinfo{year}{2009}).

\end{thebibliography}

\end{document}